# THz nanofocusing with cantilevered THz-resonant antenna tips

*Stefan Mastel[1], Mark B. Lundeberg[2], Pablo Alonso-González[1,3], Yuanda Gao[4], Kenji Watanabe[5], Takashi Taniguchi[5], James Hone[4], Frank H. L. Koppens[2,6], Alexey Y. Nikitin[1,7], Rainer Hillenbrand[*7,8]*

[1]CIC nanoGUNE, 20018 Donostia-San Sebastian, Spain

[2]ICFO-Institut de Ciències Fotòniques, The Barcelona Institute of Science and Technology, 08860 Castelldefels (Barcelona), Spain

[3]Departamento de Física, Universidad de Oviedo, 3307 Oviedo, Spain

[4]Department of Mechanical Engineering, Columbia University, New York City 10027, USA

[5]National Institute for Material Science, 1-1 Namiki, Tsukuba 305-0044, Japan

[6]ICREA – Institució Catalana de Recerça i Estudis Avancats, 08010 Barcelona, Spain

[7]IKERBASQUE, Basque Foundation for Science, 48013 Bilbao, Spain

[8]CIC nanoGUNE and UPV/EHU, 20018 Donostia-San Sebastian, Spain

*corresponding author: r.hillenbrand@nanogune.eu

We developed THz-resonant scanning probe tips, yielding strongly enhanced and nanoscale confined THz near fields at their tip apex. The tips with length in the order of the THz wavelength ($\lambda$ = 96.5 $\mu$m) were fabricated by focused ion beam (FIB) machining and attached to standard atomic force microscopy (AFM) cantilevers. Measurements of the near-field intensity at the very tip apex (25 nm radius) as a function of tip length – via graphene-based (thermoelectric) near-field detection - reveal their first and second order geometrical antenna resonances for tip length of 33 and 78 $\mu$m, respectively. On resonance, we find that the near-field intensity is enhanced by one order of magnitude compared to tips of 17 $\mu$m length (standard AFM tip length), which is corroborated by numerical simulations that further predict remarkable intensity enhancements of about $10^7$ relative to the incident field. Because of the strong field enhancement and standard AFM operation of our tips, we envision manifold and straightforward future application in scattering-type THz near-field nanoscopy and THz photocurrent nano-imaging, nanoscale nonlinear THz imaging or nanoscale control and manipulation of matter employing ultrastrong and ultrashort THz pulses.

Terahertz (THz) radiation (*1*) (*2*), loosely defined between 0.1 and 10 THz (wavelength λ = 3000 – 30 $\mu$m) (*1*), can access vibrational and rotational resonances in molecules (*3*) (*4*) (*5*) and low-energy dynamic processes in solid-state matter or devices (*4*) (*6*) (*7*). For many applications, a strong THz field concentration is required, for example, for high-resolution THz imaging or for THz sensing of small amounts of matter (*1*) (*3*). This can be accomplished by focusing THz radiation using far-field optics. However, the focal spot size is limited by diffraction to about λ/2 = 15 – 1500 $\mu$m. A nanoscale field confinement can be achieved by concentrating THz radiation with the use of metal antennas (*8*), sharp metal wires (*9*) (*10*) (*11*) (*12*) (*13*), or subwavelength apertures or slits (*12*) (*14*) (*15*) (*16*). In particular, the THz field concentration at a sharp tip apex can be achieved by exploiting the lightning rod effect, or by adiabatic compression of an electromagnetic wave propagating along a long, tapered metal wire (*10*) (*11*) (*13*) (*17*) (*18*) (*19*). Field confinements as large as λ/4600 have been already reported (*20*). Applications of near-field enhancement at nanoscale metal tips include the THz control of photoemission (*21*), nanoscale-resolved THz scattering-type scanning near-field microscopy (s-SNOM) (*22*) (*23*) (*24*) (*25*), ultrafast sub-cycle THz nano-spectroscopy (*26*) or THz photocurrent nanoscopy (*27*).

In many applications, the illuminated metal tip is much longer than the THz wavelength λ, in order to guarantee strong near-field enhancements and scattering from the tip. For subwavelength-scale THz imaging, the rather long tips of a scanning tunneling microscope (STM) (*28*) (*29*) (*30*) can be employed. In case of non-conducting samples, the long metal tips can be scanned over the sample surface via shear-force control that utilizes a tuning fork (*31*) (*32*). Alternatively, the tips of standard AFM cantilevers may be used for THz near-field imaging (*24*) (*26*). While this approach can be performed with standard and easy-to-use AFM instrumentation, the AFM tips suffer from low field enhancement due to the large mismatch between tip length (<< λ) and THz wavelength λ. AFM tips of a length in the order of the THz wavelengths - potentially exhibiting geometric antenna resonances that provide large field enhancements - have not been developed yet, despite their advantage to enable nanoscale THz control and imaging applications based on widely available AFM instrumentation.

Here, we developed cantilevered antenna probes with nanoscale tip apex for resonant nanofocusing of THz radiation. Their lengths were designed to support antenna modes to resonantly enhance the THz field at the tip apex. We attached the antennas to standard atomic force microscopy (AFM) cantilevers to allow for a precise control of the position of the THz hotspot on a sample surface using standard AFM instrumentation. To characterize the antenna probes, we measured the near field intensity directly at the tip apex using a graphene-based THz photodetector (*27*) (*33*), rather than deducing it by detecting the tip-scattered light in the far field. We find that our tips support antenna resonances and corroborate our findings with numerical simulations and antenna theory.

Fig. 1a shows a false color scanning electron microscopy (SEM) image of a FIB fabricated THz antenna probe using a Helios 450 DualBeam (FEI, Netherlands) electron microscope (*34*) (*35*). A detailed description of the fabrication process is given in the supplement. We used standard Si AFM cantilevers (Nanoworld, Switzerland) and replaced the original tip by a several tens of micrometers long tip made of an 80/20 Pt/Ir alloy. To achieve a high field confinement and enhancement, the tip apex diameter is adjusted to only (50 +/- 3) nm. We fabricated six different tips with lengths 17$\mu$m, 33$\mu$m, 43$\mu$m, 55$\mu$m, 65$\mu$m, and 78$\mu$m, each of which supports a different antenna mode at one given excitation THz wavelength.

To characterize the cantilevered THz antennas, we employed them as scanning probe tips in a scattering-type Scanning Near-field Optical Microscope (s-SNOM, Neaspec GmbH, Germany). The s-SNOM is based on a non-contact atomic force microscope (AFM), where the tip is oscillating vertically at the mechanical resonance frequency $\Omega$ of the cantilever. In the present work, the oscillation amplitude was 40 nm. The tips were illuminated with the focused THz beam ($\lambda$ = 96.5 $\mu$m, 3.11 THz) of a gas laser (SIFIR-50 FPL, Coherent Inc., USA), which provides monochromatic radiation up to 100 mW power. In contrast to standard s-SNOM, we did not detect the tip-scattered field but used a graphene-based THz detector (_27_) (_36_) (illustrated in Fig. 1b and described in the Methods section) to measure the near-field intensity directly at the tip apex. The detector, in brief, consists of a graphene sheet encapsulated in two hexagonal Boron Nitrite (h-BN) layers on top of two laterally separated gates $G_L$ and $G_R$. By applying two different gate voltages $V_L$ and $V_R$, we generated a pn-junction in the graphene across the gap between the two gates. The near fields at the tip apex locally heat the electrons in the graphene, which induces close to the junction a thermoelectric photocurrent (_27_) (_36_) (_37_) (_38_). This photocurrent can be measured through the two lateral contacts $C_L$ and $C_R$, and is found to be proportional to the near-field intensity for the power applied in our experiments, as shown in the inset in Fig. 2b (see also supplement). We note that the direct detection of the tips' near field offers the advantage that only the tip illumination needs to be adjusted. There is no need for a detection beam path, which typically comprises an interferometer (*39*) that requires not only accurate adjustment of the collection and detector optics, but also of the beam quality and wavefronts. This significant reduction of adjustment steps enables a more reliable and accurate comparison of the near-field enhancement at the apex of various different tips.

We first demonstrate that the antenna probes allow for stable AFM imaging and nanoscale THz focusing. To that end, we recorded a topography image (Fig. 1c) of the detector device (using the 78 $\mu$m long antenna probe), showing the top surface (h-BN layer) of the detector above the active region, as well as the lateral Au contacts (left and right) collecting the photocurrent. It clearly verifies a stable AFM operation using our THz antenna probes, despite their comparably large size and hence mass (~60 pg > 80 times the mass of standard Si tip), which reduces the mechanical cantilever resonance frequency by nearly a factor two (from 252 kHz for cantilever with standard Si tip to 139 kHz with THz antenna probe of length 78 $\mu$m). To demonstrate the THz nanofocusing functionality of the antenna probe, we recorded a DC photocurrent image $I_{PC,DC}$ (Fig. 1d) simultaneously to the topography. We see a bright vertical stripe of strong photocurrent $I_{PC,DC}$ in the image center, which reveals the strong photo-thermoelectric current generation near the pn-junction. The stripe has a sub-wavelength full width at half maximum of ~ 0.6 $\mu$m, which verifies that the THz radiation can be focused by the tip to a deeply subwavelength scale spot. Further, we observe a strong photocurrent $I_{PC,DC}$ close to the lateral source and drain contacts. It arises from a less-defined local doping of the graphene near the contacts (*40*) (*41*). The photocurrent abruptly drops to a constant background value (see discussion in following paragraph) at the graphene edge (marked by the white dashed line in Fig.

1d) and at the metal contacts. From the signal change at the contact we estimate spatial resolution (*i.e.* lateral field confinement at the tip apex) of about 100 nm ($\lambda/1000$), verifying the conversion of incoming THz radiation into a highly confined nanofocus at the tip apex, and hence the functionality of our tips as high-resolution THz near-field probes.

For quantifying the vertical field confinement, we recorded the photocurrent $I_{PC,DC}$ as a function of distance *h* between tip and detector (solid red curve in Fig. 2a) at the position marked by a black cross in Fig. 1e. The photocurrent $I_{PC,DC}$ decays rapidly with increasing *h*. For large *h* it approaches asymptotically the constant value of 3.3 nA, which we assign to a background photocurrent $I_{PC,BG}$ that is generated by the diffraction-limited illumination of the whole device. Knowing $I_{PC,BG}$, we can extract the near-field contribution $\Delta I_{PC} = I_{PC,DC} - I_{PC,BG}$ to determine vertical confinement (1/e decay length *d*) of the THz near field (Fig. 2b). We measure $d = 28$ nm, revealing a deep subwavelength-scale vertical field confinement at the tip apex (amounting to about $\lambda/3500$), which agrees well with the numerically calculated near-field distribution at the tip apex (50 nm diameter) of a 78 μm long Pt tip (inset of Fig. 2b).

Interestingly, the background contribution ($I_{PC,BG} = 3.3$ nA) is remarkably small compared to the near-field signal, $\Delta I_{PC} = 15.1$ nA, which typically is not the case in scattering-type and tip-enhanced near-field techniques. We explain the finding by the small active area of the THz detector, which is significantly smaller than the THz focus illuminating the tip. The small but non-negligible background signal can be fully suppressed by demodulating the detector signal at harmonics $n\Omega$ of the tip oscillation frequency $\Omega$ (similar to s-SNOM and infrared photocurrent nanoscopy (*24*) (*38*) (*42*)), yielding the signal $I_{PC,n\Omega}$. Recording $I_{PC,n\Omega}$ as a function of tip-detector distance $h$ for n = 1 and 2 (dashed red curves in Fig. 2a) indeed shows that the demodulated photocurrent signal completely vanishes for large tip-detector distances $h$. Due to the "virtual tip-sharpening" effect by higher harmonic signal demodulation (*43*) (*44*), we measure a decreasing 1/e decay length of $d_1$ = 17 nm ($\lambda$/5600) and $d_2$ = 9 nm ($\lambda$/10500) for n = 1 and n = 2, respectively. The demodulation also allows for background-free photocurrent nanoimaging, as demonstrated in Fig. 1e (demodulation at n = 1), where the photocurrent drops to $I_{PC,n\Omega}$ = 0 nA on the lateral Au contacts and on the $SiO_2$ substrate (white areas in Fig. 1e).

Having verified a proper AFM operation and near-field focusing performance of the FIB-fabricated tips, we compare in the following the near-field intensity at the apex of differently long tips. In Fig. 2b we compare $\Delta I_{PC}$ as a function of tip-detector distance $h$ for a 78 $\mu$m and a 17 $\mu$m long tip. The measurements were taken at the same position on the photodetector, marked by a black cross in Fig. 1d (5.5 $\mu$m from the device edge along the pn-junction). While the background corrected signal $\Delta I_{PC}$ at large distances $h$ converges to zero for both tips, we observe at contact ($h$ = 0 nm) a significantly enhanced photocurrent for the 78 $\mu$m long tip. For more detailed insights into the dependence of the near-field intensity enhancement on the tip length, we performed photocurrent measurements with six differently long tips. To that end, we recorded line profiles of $\Delta I_{PC}$ (average of 100, marked in Fig. 1d by dashed black horizontal line) across the pn-junction. The recording of line profiles, rather than approach curves, offers the advantage that measurement errors due to uncertainties in tip positioning can be minimized. Note that we did not analyze the background-free demodulated photocurrent signals $I_{PC,n\Omega}$, since they do not reveal the near-field intensity but the vertical gradients of the near-field intensity. In Fig. 3a we plot three line profiles showing the near-field photocurrent $\Delta I_{PC}$ obtained with tips of length L = 17 $\mu$m, 33 $\mu$m, and 78 $\mu$m. All three curves exhibit a maximum near-field photocurrent $\Delta I_{PC,max}$ at the position of the pn-junction (x = 0 nm), and decay to either side towards the source and drain contacts. As seen before in Fig. 2b, we find a strong variation of the near-field photocurrent for the different tips. Plotting $\Delta I_{PC,max}$ as a function of antenna length L for the six different tips (blue dots in Fig. 3b), we find that $\Delta I_{PC,max}$ strongly depends on the tip length L, indicating minima and maxima and thus antenna resonances. The longest antenna probe (L = 78$\mu$m) yields the strongest, nearly nine-fold near-field intensity enhancement compared to the shortest tip (L = 17 $\mu$m). Note that both the tip length and the tip apex diameter determine the photocurrent signal. A larger tip diameter reduces the lateral field confinement below the tip, thus illuminating the detector on a larger area, while the field enhancement is reduced. For a constant tip diameter it can be shown that a variation of the tip length only varies the field enhancement but not the field confinement (see supplement S4). Hence, we can isolate the effect of the antenna length (field enhancement) on the photocurrent by adjusting the apex diameter for each tip to a constant value. For the presented experiments, we fabricated tips with a diameter of 50 nm, which was highly reproducible with an accuracy of +/- 3 nm.

To elucidate the variations of the near-field enhancement for different tips, we performed numerical full-wave simulations (see Methods) of tips, illuminated with THz radiation, with a geometry as depicted in Fig. 3c (for more detail see schematics D in Fig. 4a). We assume a p-polarized plane wave illumination (electric field $E_{inc}$) at 3.11 THz ($\lambda$ = 96.5 µm) at an angle of $\alpha$ = 60° relative to the tip axis. The tip (with small Si cantilever attached at its shaft) is placed $h$ = 20 nm above the surface of a detector consisting of a 9 nm thick hBN layer that covers a graphene layer on top of a bulk hBN substrate. The blue curve in Fig. 3b shows the calculated near-field intensity enhancement $f = \left(\frac{E_{nf}}{E_{in}}\right)^2$ between tip and hBN surface (10 nm below the tip). An excellent agreement with the experimentally measured near-field photocurrent (blue dots) is observed. Particularly, the calculation exhibits the maxima at tip lengths of about $L_{res,1} = 34$ $\mu$m and $L_{res,2} = 81$ $\mu$m. The logarithm of the near-field distributions shown in Fig. 3d let us identify the maxima as first and second order antenna resonance, respectively. The latter is excited because of retardation along the tip axis, caused by the inclined illumination relative to the tip axis (45). The two resonances yield an impressive field intensity enhancement of about $1.2 \times 10^7$ and $2 \times 10^7$. Most important, the resonant tips increase the field intensity enhancement by about one order of magnitude compared to the 17 $\mu$m long tip, which length is that of standard AFM tips.

Compared to classical dipolar radio wave antennas (45) – where $L_{res,n} = n\lambda/2$ with $n$ being the resonance order – we find that i) the antenna tip's resonances occur at shorter lengths, and ii) their resonance lengths do not scale linearly with $n$ (we measure $L_{res,1} = \lambda/2.82$ and $L_{res,2} = \lambda/1.19$) These deviations may be explained by resonance shifts caused by the presence of the cantilever and/or photodetector. To understand the resonance shifts and to establish future design rules for resonant THz probes, we performed simulations considering a systematic variation of the tip's environment. Fist, we calculated the near-field intensity enhancement 10 nm below the apex of an isolated antenna tip (illustrated by sketch A in in Fig. 4a) as a function of the tip length (black curve, Fig. 4b). In good agreement with classical antenna theory (45) ($L_{res,n} = n\lambda/2$), we find the first two antenna resonances at $L_{res,1} = 44$ $\mu$m $= \lambda/2.19$ and $L_{res,2} = 89$ $\mu$m $= \lambda/1.08$. The small deviation from $L_{res,n} = n\lambda/2$ we explain by the conical shape of the tip (45). By adding a silicon cantilever to the tip shaft (sketch B in Fig. 4a), the resonance length of the calculated spectrum (red curve in Fig. 4b) shift to $L_{res,1} = 34$ $\mu$m $= \lambda/2.8$ and $L_{res,2} = 81 \mu$m $= \lambda/1.2$, while the peak height is reduced by about 27 and 17 percent, respectively. Both observations can be explained by a capacitive loading of the tip antenna by the Si cantilever (*45*). Next, the sample (detector device) is considered in the simulations (sketch C in Fig. 4a). It is placed 20 nm below the tip apex, and the field enhancement is measured 10 nm below the tip. A detailed description of the simulation parameters is given in the methods section. The calculated spectrum is shown by the blue curve in Fig. 4b. Compared to geometry B (red curve in Fig. 4b), the near-field intensities at the resonance lengths $L_{res,1}$ and $L_{res,2}$ are significantly enhanced by a factor of about seven. This enhancement can be explained by the near-field coupling between tip and sample. Interestingly, the near-field coupling does not further shift the antenna resonance, which typically occurs at visible and infrared frequencies when an antenna is brought in close proximity to a dielectric or metallic sample (*46*).

To better understand the absence of resonance shifts due to tip-sample coupling, we first studied the role of the graphene in the near-field coupling. We repeated the numerical calculation, but replaced the graphene with a perfect electric conductor (PEC) (geometry D in Fig. 4a). Although the PEC perfectly screens the near fields at the tip apex, the antenna spectrum (gray curve, Fig. 4b) shows only a minor increase of the peak heights of about twenty percent, and a minor resonance length shift ($L_{res,1} = 33.5$ $\mu$m $= \lambda/2.9$ and $L_{res,2} = 80.5$ $\mu$m $= \lambda/1.2$) compared to geometry C (blue curve, Fig. 4b). The results imply that graphene at THz frequencies acts as a nearly metallic reflector for the tip's near fields. The results imply that graphene at THz frequencies acts as a nearly metallic reflector for the tip's near fields. This can be explained by the convergence of the Fresnel reflection coefficient towards one for the large wavevectors associated with the near fields at the tip apex (*47*). Consequently, strong near-field coupling between tip and graphene occurs, leading to strongly enhanced field at the tip apex. In this regard, the nearly negligible spectral shift of the antenna resonance may be even more surprising.

We explain the negligible spectral shift with the help of radio frequency (RF) theory (*45*). In the RF range, circuit theory is an essential tool for the efficient design of antennas, and has recently been adopted for the visible and infrared spectral range (*48*) (*49*) (*50*). We consider the tip above the sample as an antenna arm (for simplicity a thin metal rod) above a metallic ground plane. A sketch and the corresponding circuit model are shown in Figs. 5a and b. The antenna arm (rod above) is described by its intrinsic (dipole) impedance, $Z_A = R_A + i X_A$, where $R_A$ and $X_A$ are the dipole's resistance and reactance, respectively (see Fig. 5d) (*45*). The air gap between tip and sample can be considered as a capacitive load with impedance given by (*48*)

$$Z_{gap} = \text{R}_{gap} + i\,\text{X}_{gap} = -\frac{ih}{\omega \varepsilon D^2} \qquad (1)$$

where $h$ is the gap height, $\omega$ the THz frequency, $\varepsilon = 1$ (air) the dielectric permittivity of the gap filling medium and $D$ the diameter of both the antenna arm and the gap. Because of the open circuit operation of our antenna (the antenna is neither connected to a source nor a receiver), the input impedance $Z_{in} = R_{in} + i X_{in}$ of the antenna can be considered as a serial combination of the two impedances $Z_A$ and $Z_{gap}$ (*49*) (*50*) (*51*). In this circuit model (Fig. 5a and b), a resonance occurs when $X_{in} = 0$ (*48*) (*52*), *i.e.* when the capacitive reactance of the load cancels the intrinsic inductive reactance of the antenna, $-X_{gap} = X_A$.

To understand the antenna resonance, we discuss $X_A$ and $X_{gap}$ as a function of the antenna arm length L. The red curve in Fig. 5c shows $X_A$ for an illumination wavelength $\lambda = 96.5$ $\mu$m. It was calculated according to reference (*45*) (see Methods), assuming a metal rod of diameter D = 50 nm (corresponding to the tip apex diameter). We find $X_A = 0$ for $L \approx \lambda/4$, which represents the first closed circuit resonance of a classical RF antenna comprising a metal rod (of length L) on a ground plane, not considering the air gap yet. At $L \approx \lambda/2$ we find that $X_A$ diverges, indicating the first open circuit (scattering) resonance (*48*). To see how the antenna resonance depends on the capacitive coupling across the air gap, we plot the capacitive reactance $|X_{gap}|$ for gap heights of $h$ = 4 nm and 5 nm (horizontal dashed red lines in Fig. 5c). We observe that $|X_{gap}|$ decreases with decreasing gap height (*i.e.* the gap capacitance increases) and the intersection between $X_A$ and $|X_{gap}|$ (resonance condition) shifts the antenna resonance length $L_{res}$ from $\lambda/2$ towards $\lambda/4$ for further decreasing gap width (see also Fig. 5d). Interestingly, the resonance length $L_{res} \approx \lambda/2$ barely shifts until gap heights as small as 5 nm are reached. Obviously, the capacitance of an air gap larger than 5 nm is negligible small and thus yields a large capacitive reactance that is comparable to that of the antenna close to its open circuit resonance.

We show in Figure 5d the antenna resonance length $L_{res}$ as a function of the gap width $h$ (red curve). For $h > 5$ nm we find that $L_{res}$ is nearly constant and only slightly smaller than $\lambda/2$. Only in close proximity to the substrate (h < 5 nm) the resonance length rapidly decreases. For comparison, we numerically calculated the antenna resonance length of a metal tip above a perfectly conducting ground plane. The result (inset Fig. 5d) confirms that the antenna resonance of a tip does not shift for tip-sample distances larger than 5 nm, although the antenna resonance length ($L_{res}$ = 44 $\mu$m = $\lambda$ / 2.19) is slightly smaller than that obtained by antenna theory (which can be attributed to the conical shape of the tip, which is not considered in our antenna circuit model). Based on these theoretical results we can explain the absence of resonance shifts in our experiments and the numerical simulations shown in Figs. 3 and 4 by the relatively large average distance $h$ = 30 nm between tip and graphene. We conclude that in future design of THz resonant probes and interpretation of results one needs to consider the possibility of resonance shifts only for very small tip-sample distances depending on tip radius.

We finally discuss our results in the wider context of optical antennas. We used the antenna circuit model to calculate the resonance shifts for a mid-infrared illumination wavelength ($\lambda$ = 9.6 $\mu$m; gray curve in Fig. 5d. For the same antenna diameter D, we observe that a significant shift of the resonance length $L_{res}$ occurs already at much larger gap width h. This can be attributed to the decreasing capacitive gap reactance $X_{gap}$ when the frequency is increased (Eq. 1), while the inductive antenna reactance $X_A$ barely changes (compare grey and red curves in Fig. 5c). We note that our calculations do not consider plasmonic effects, which at higher frequencies cause further resonance shifts, although not being the root cause for them.

In summary, we have demonstrated the FIB fabrication of sharp, several tens of micrometer long THz antenna tips on standard AFM cantilevers. To evaluate their performance, we applied a graphene-based THz detector to measure the relative near-field intensity directly at the tip apex. The tips were found to support strong antenna resonances, in excellent agreement with numerical calculations. At resonance, the tips provide a nine-fold near-field intensity enhancement at the tip apex as compared to tips of a length that is typical in AFM, while the numerical simulations predict resonant near-field intensity enhancement factors of up to $10^7$ relative to the incident field. Our nanoscale THz-resonant near-field probes promise exiting future applications, including scattering-type THz near-field microscopy with enhanced sensitivity, nanoscale nonlinear THz imaging or nanoscale control and manipulation of matter using ultrastrong and ultrashort THz pulses (*53*) (*54*) (*7*) (*55*) (*56*). We envision even stronger field enhancement by further reducing the tip apex diameter form currently 50 nm to well below 10 nm.

METHODS

**Split-gate graphene detector**

The detector (*27*) (*36*) consists of a graphene sheet encapsulated between two layers (9nm top, 27nm bottom) of hexagonal Boron Nitride (hBN). This hBN-graphene-hBN heterostructure is placed on top of two gold backgates, which are laterally separated by a gap of 150 nm. By applying voltages $V_L$ and $V_R$ to the gates, the carrier concentration in the graphene can be controlled separately. In our experiment we have chosen the carrier concentrations $n_{L/R} = +/- 2.6 \times 10^{11}$ cm$^{-1}$, yielding a sharp pn-junction across the gap between the two gates. When the tip is placed above the gap, the near field at the apex locally heats the electrons in the graphene, yielding a photocurrent $I_{PC}$ according to $I_{PC} = (S_L - S_R) * \Delta T$ (*27*) (*57*) (*37*). Here, $\Delta T$ is the local temperature gradient below the tip and $(S_L - S_R)$ is the local variation of the Seebeck coefficient $S$ (in our device generated by the strong carrier density gradient *i.e.* the pn-junction above the gap). The photocurrent $I_{PC}$ is measured via the two lateral source and drain gold contacts. The detector is operated in its linear regime (*58*) for the power applied in the experiments, as shown in Fig.

2b. Then, for fixed gate voltages, the photocurrent $I_{PC}$ is proportional to the temperature gradient, which in turn is proportional to the near-field intensity at the tip apex (*38*).

**Fourier Filtering of DC photocurrent signals**

During the measurement of the DC approach curves (Fig. 2b) and the line profiles (Fig. 3 a) a periodic noise of 50 Hz could not fully be eliminated. To correct the data we used Fourier analysis, where first the respective data set was Fourier transformed. In Fourier domain we identified the frequency $f_0$ corresponding to 50 Hz and removed the respective data points. The removed data points were replaced by a linear interpolation between the two adjacent points. Finally, the inverse Fourier transformation of the resulting data set yields the presented DC approach curves (Fig. 2b) and line profiles (Fig. 3a). To illustrate the effect of the filtering procedure, we show in the supplementary Fig. S4 one filtered line profile in comparison with the original data.

**Numerical Simulations**

The numerical simulations were conducted using the commercial software Comsol (www.comsol.com, Stockholm, Sweden) based on finite element methods in the frequency domain. In all simulations, the conical tip had an apex radius R = 20 nm and a ratio length/width=8, which in good approximation represents the experimentally fabricated tips. For the metal we used a dielectric permittivity of Pt $\epsilon_{Pt} = -5500 + i*12000$ resulting from a Drude model fit in reference (*59*). The part of the cantilever, to which the tips were attached, was simulated as a piece of silicon of 6 $\mu$m thickness (obtained from SEM image) and 5 $\mu$m length and width. The length and width were chosen to obtain convergence of the numerical simulations. The tip was illuminated by a plane wave $E_{in}$ with wavelength $\lambda = 96.5 \mu m$ (3.11 THz) at an angle of 60° relative to the tips axis. The sample was placed 20 nm beneath the tip, while the electric field enhancement $E_{nf}$ was calculated 10 nm below the tip apex.

We simulated the graphene with a Fermi energy $E_F = v_f * \hbar * \sqrt{|n| * \pi} \approx 300 \: meV$, a relaxation time $\tau = \frac{\mu E_F}{v_f^2} \approx 1.2 \: ps$, with Fermi velocity $v_F = 10^8 \: cm \: s^{-1}$, and carrier sheet density $n = 6.57 * 10^{12} \: cm^{-2}$. We assumed high quality graphene with a mobility of $\mu = 40000 \: cm^2/V * s$ ($60$). The gate voltages were converted to carrier sheet densities via $n_{L,R}$=(0.73 x $10^{16}$ $m^{-2}$ $V^{-1}$)($V_{L,R}$-$V_{CNP}$). $V_{CNP}$=0.15V is the gate voltage at the charge neutrality point (CNP), which was determined by examining the gate dependence of the device. The coefficient $0.73 \times 10^{16} \: m^{-2} V^{-1}$ was calculated as the static capacitance of the 27 nm thick hBN bottom layer with dielectric constant 3.56 ($37$).

**Antenna Theory**

The antenna impedance $Z_A = R_A + i * X_A$ was calculated using standard equations from RF antenna theory ($45$). The antenna resistance $R_A$ (neglecting ohmic losses) and reactance $X_A$ are given by

$$R_A = \frac{1}{2} \frac{\eta}{2\pi \sin\left(\frac{kl}{2}\right)^2} (C + \ln(kl) - C_i(kl) + \frac{1}{2}\sin(kl)\left(S_i(2kl) - 2S_i(kl)\right) + \frac{1}{2}\cos(kl)(C$$
$$+ \ln\left(\frac{kl}{2}\right) + C_i(2kl) - 2C_i(kl)))$$

and

$$X_A = \frac{1}{2} \frac{\eta}{4\pi \sin\left(\frac{kl}{2}\right)^2} (2S_i(kl)\cos(kl)(2S_i(kl) - S_i(2kl)) - \sin(kl)(2C_i(kl) - C_i(2kl))$$

$$- C_i\left(\frac{kD^2}{2l}\right)))$$

where $C = 0.5772$ is the Euler constant, $k$ is the wave vector of the electromagnetic wave, $l$ is the antenna length, $D$ is the antenna diameter, $\eta$ is the impedance of the surrounding medium (for free space $\eta = 377\Omega$) and $S_i$ and $C_i$ are the sine and cosine integrals given by $S_i(z) = \int_z^\infty \frac{\sin(t)}{t} dt$ and $C_i(z) = \int_z^\infty \frac{\cos(t)}{t} dt$.

## ASSOCIATED CONTENT

**Supporting Information**. The following files are available free of charge in the supplement (PDF).

- A detailed description of the fabrication process of the THz antenna tips used in this work.

- Scanning electron microscopy (SEM) images to measure the length of the probes.

- A description of the measurement of the linearity of the detector device.

- A Description of the Fourier filtering of the DC photocurrent signals.

- Numerical Calculation of the filed confinement below the tip apex.

## AUTHOR INFORMATION

**Corresponding Author**

*Email: r.hillenbrand@nanogune.eu

**Notes**

The authors declare the following competing financial interest (s): R. Hillenbrand is co-founder of Neaspec GmbH, a company producing scattering-type scanning near-field optical microscopy systems such as the one used in this study. All other authors declare no competing financial interests.

ACKNOWLEDGMENT

The authors thank Andrey Chuvilin for sharing his ideas and knowledge about the fabrication methodology of very long cantilevered AFM probes. Further, we thank Nader Engheta and Martin Schnell for fruitful discussions about antenna theory. The authors acknowledge financial support from the European Commission under the Graphene Flagship (contract no. CNECT-ICT-604391) and the Spanish Ministry of Economy and Competitiveness (national project MAT2015-65525-R and MAT2014-53432-C5-4-R).

REFERENCES

1. Ferguson, B.; Zhang, X.-C. *Nature Materials* **2002,** *1* (1), 26-33.

2. Tonouchi, M. *Nature Photonics* **2007,** *1*, 97-105.

3. Mittleman, D., Ed. *Sensing with Terahertz Radiation;* Springer-Verlag: Berlin, 2003.

4. Baxter, J. B.; Guglietta, G. W. *Analytical Chemistry* **2011,** *83*, 4342-4368.

5. Jepsen, P. U.; Cooke, D. G.; Koch, M. *Laser & Photonics Review* **2011,** *5* (1), 124-166.

6. Ulbricht, R.; Hendry, E.; Shan, J.; Heinz, T. F.; Bonn, M. *Reviews of Modern Physics* **2011,** *83* (2), 543-586.

7. Kampfrath, T.; Tanaka, K.; Nelson, K. A. *Nature Photonics* **2013,** *7* (9), 680-690.

8. Toma, A.; Tuccio, S.; Prato, M.; De Donato, F.; Perucchi, A.; Di Pietro, P.; Marras, S.; Liberale, C.; Proietti Zaccaria, R.; De Angelis, F.; et al. *Nano Letters* **2015,** *15* (386-391).

9. van der Valk, N. C. J.; Planken, P. C. M. *Applied Physics Letters* **2002,** *81* (9), 1558-1560.

10. Maier, S. A.; Andrews, S. R.; Martín-Moreno, L.; García-Vidal, F. J. *Physical Review Letters* **2006,** *97* (17), 176805.

11. Astley, V.; Rajind, M.; Mittleman, D. M. *Applied Physics Letters* **2009,** *95* (3), 031104.

12. Adam, A. J. L. *Journal of Infrared Millimeter and Terahertz Waves* **2011,** *32*, 976-1019.

13. Mittleman, D. M. *Nature Photonics* **2013,** *7*, 666-669.

14. Hunsche, S.; Koch, M.; Brenner, I.; Nuss, M. C. *Optics Communications* **1998,** *150*, 22-26.

15. Seo, M. A.; Park, H. R.; Koo, S. M.; Park, D. J.; Kang, J. H.; Suwal, O. K.; Choi, S. S.; Planken, P. C. M.; Park, G. S.; Park, N. K.; et al. *Nature Photonics* **2009,** *3* (3), 152-156.

16. Zhan, H.; Mendis, R.; Mittleman, D. M. *Optics Express* **2010,** *18* (9), 9643-9650.

17. Liang, H.; Ruan, S.; Zhang, M. *Optics Express* **2008,** *16* (22), 18241-18248.

18. Wang, K.; Mittleman, D. M. *Nature* **2004,** *432*, 376-379.

19. Awad, M.; Nagel, M.; Kurz, H. *Applied Physics Letters* **2009,** *94* (5), 051107.

20. Kuschewski, F.; von Ribbeck, H.-G.; Döring, J.; Winnerl, S.; Eng, L. M.; Kehr, S. C. *Applied Physics Letters* **2016,** *108*, 113102.

21. Wimmer, L.; Herink, G.; Solli, D. R.; Yalunin, S. V.; Echterkamp, K. E.; Ropers, C. *Nature Physics* **2014,** *10*, 432-436.

22. Chen, H.-T.; Cho, G. C.; Kersting, R. *Applied Physics Letters* **2003,** *83* (15), 3009.

23. von Ribbeck, H.-G.; Brehm, M.; van der Weide, D. W.; Winnerl, S.; Drachenko, O.; Helm, M.; Keilmann, F. *Optics Express* **2008,** *16* (5), 3430-3438.

24. Huber, A. J.; Keilmann, F.; Wittborn, J.; Aizpurua, J.; Hillenbrand, R. *Nano Letters* **2008,** *8* (11), 3766-3770.

25. Moon, K.; Park, H.; Kim, J.; Do, Y.; Lee, S.; Lee, G.; Kang, H.; Han, H. *Nano Letters* **2012,** *15*, 549-552.

26. Eisele, M.; Cocker, T. L.; Huber, M. A.; Plankl, M.; Viti, L.; Ercolani, D.; Sorba, L.; Vitiello, M. S.; Huber, R. *Nature Photonics* **2014,** *8,* 841-845.

27. Alonso-González, P.; Nikitin, A. Y.; Gao, Y.; Woessner, A.; Lundeberg, M. B.; Principi, A.; Forcellini, N.; Yan, W.; Vélez, S.; Huber, A. J.; *et al*. *Nature Nanotechnology* **2017,** *12,* 31-35.

28. Cocker, T. L.; Jelic, V.; Gupta, M.; Molesky, S. J.; Burgess, J. A. J.; Reyes, G. D. L.; Titova, L. V.; Tsui, Y. Y.; Freeman, M. R.; Hegmann, F. A. *Nature Photonics* **2013,** *7* (8), 620-625.

29. Cocker, T. L.; Peller, D.; Yu, P.; Repp, J.; Huber, R. *Nature* **2016,** *539,* 263-267.

30. Jelic, V.; Iwaszczuk, K.; Nguyen, P. H.; Raghje, C.; Hornig, G. J.; Sharum, H. M.; Hoffman, J. R.; Freeman, M. R.; Hegmann, F. A. *Nature Physics* **2017**.

31. Giessibl, F. J. *Applied Physics Letters* **1998,** *73* (26), 3956-3958.

32. Buersgens, F.; Acuna, G.; Lang, C. H.; Potrebic, S. I.; Manus, S.; Kersting, R. *Review of Scientific Instruments* **2007,** *78* (113701), 113701.

33. Cai, X.; Sushkov, A. B.; Suess, R. J.; Jadidi, M. M.; Jenkins, G. S.; Nyakiti, L. O.; Myers-Ward, R. L.; Li, S.; Yan, J.; Gaskill, D. K.; *et al*. *Nature Nanotechnology* **2014,** *9,* 814-819.

34. Huth, F.; Chuvilin, A.; Schnell, M.; Amenabar, I.; Krutokhvostov, R.; Lopatin, S.; Hillenbrand, R. *Nano Letters* **2013,** *13,* 1065-1072.

35. Wang, A.; Butte, M. J. *Applied Physics Letters* **2014,** *105* (053101).

36. Lundeberg, M. B.; Gao, Y.; Woessner, A.; Tan, C.; Alonso-González, P.; Watanabe, K.; Taniguchi, T.; Hone, J.; Hillenbrand, R.; Koppens, F. H. L. *Nature Materials* **2016,** *14,* 421-425.

37. Wössner, A.; Lundeberg, M. B.; Gao, Y.; Principi, A.; Alonso-González, P.; Carrega, M.; Watanabe, K.; Taniguchi, T.; Vignale, G.; Polini, M.; *et al*. *Nature Materials* **2014,** *14,* 421-425.

38. Wössner, A.; Alonso-González, P.; Lundeber, M. B.; Gao, Y.; Barrios-Vargas, J. E.; Navickaite, G.; Ma, Q.; Janner, D.; Watanabe, K.; Cummings, A. W.; *et al*. *Nature Communications* **2016,** *7,* 10783.

39. Keilmann, F.; Hillenbrand, R. IV: Apertureless Near-Field Optical Microscopy in *Nano-optics and near-field optical microscopy, 1st edition*; Zayats, A. V.; Richards, D., Eds.; *Artech House: Boston/London,* 2009.

40. Lee, E. J. H.; Balasubramanian, K.; Weitz, R. T.; Burghard, M.; Kern, K. *Nature Nanotechnology* **2008,** *3,* 486-490.

41. Giovannetti, G.; Khomyakov, P. A.; Brocks, G.; Karpan, V. M.; van den Brink, J.; Kelly, P. J. Doping *Physical Review Letters* **2008,** *101,* 026803.

42. Ocelic, N.; Huber, A.; Hillenbrand, R. *Applied Physics Letters* **2006,** *89* (101124).

43. Knoll, B.; Keilmann, F. *Optics Communications* **2000,** *182,* 321-328.

44. Amarie, S.; Zaslansky, P.; Kajihara, Y.; Griesshaber, E.; Schmahl, W. W.; Keilmann, F. *Beilstein Journal of Nanotechnology* **2012,** *3,* 312-323.

45. Balanis, C. A. *Antenna Theory;* John Wiley & Sons: Hoboken, New Jersey, USA, 2005.

46. Novotny, L.; Hecht, B. *Principles of Nano-Optics,* 2nd ed.; Cambridge University Press: Cambridge UK, 2012.

47. Kim, D.-S.; Kwon, H.; Nikitin, A. Y.; Ahn, S.; Martin-Moreno, L.; Garcia-Vidal, F. J.; Ryu, S.; Min, H.; Kim, Z. H. *ACS Nano* **2015,** *9* (7), 6965-6773.

48. Alú, A.; Engheta, N. *Nature Photonics* **2008,** *2,* 307-310.

49. Alú, A.; Engheta, N. *Physical Review Letters* **2008,** *101* (4), 043901.

50. Engheta, N.; Salandrino, A.; Alú, A. *Physical Review Letters* **2005,** *95* (9), 095504.

51. Schnell, M.; García-Etxarri, A.; Huber, A. J.; Crozier, K.; Aizpurua, J.; Hillenbrand, R. *Nature Photonics* **2009,** *3,* 287-291.

52. Benz, F.; de Nijs, B.; Tserkezis, C.; Chikkaraddy, R.; Sigle, D. O.; Pukenas, L.; Evans, S. D.; Aizpurua, J.; Baumberg, J. J. *Optics Express* **2015,** *23* (26), 33255-33269.

53. Cole, B. E.; Williams, J. B.; King, B. T.; Sherwin, M. S.; Stanley, C. R. *Nature* **2001,** *410,* 60-63.

54. Kampfrath, T.; Sell, A.; Klatt, G.; Pashkin, A.; Mährlein, S.; Dekorsy, T.; Wolf, M.; Fiebig, M.; Leitenstorfer, A.; Huber, R. *Nature Photonics* **2010,** *5* (1), 31-34.

55. Danielson, J. R.; Lee, Y.-S.; Prineas, J. P.; Steiner, J. T.; Kira, M.; Koch, S. W. *Physical Review Letters* **2007,** *99* (237401).

56. Schubert, O.; Hohenleutner, M.; Langer, F.; Urbanek, B.; Lange, C.; Huttner, U.; Golde, D.; Meier, T.; Kira, M.; Koch, S. W.; et al. *Nature Physics* **2014,** *8,* 119-123.

57. Gabor, N. M.; Song, J. C. W.; Ma, Q.; Nair, N. L.; Taychatanapat, T.; Watanabe, K.; Taniguchi, T.; Livitov, L. S.; Jarillo-Herero, P. *Science* **2011,** *334,* 648-652.

58. Xu, X.; Gabor, N. M.; Alden, J. S.; von der Zande, A. M.; McEuen, P. L. *Nano Letters* **2010,**

*10,* 562-566.

59. Ordal, M. A.; Bell, R. J.; Alexander, R. W.; Long, L. L.; Querry, M. R. *Applied Optics* **1985,** *24* (24), 4493-4499.

60. Wang, L.; Meric, I.; Huang, P. Y.; Gao, Q.; Gao, Y.; Tran, H.; Taniguchi, T.; Watanabe, K.; Campos, L. M.; Muller, D. A.; *et al*. *Science* **2013,** *342,* 614-617.

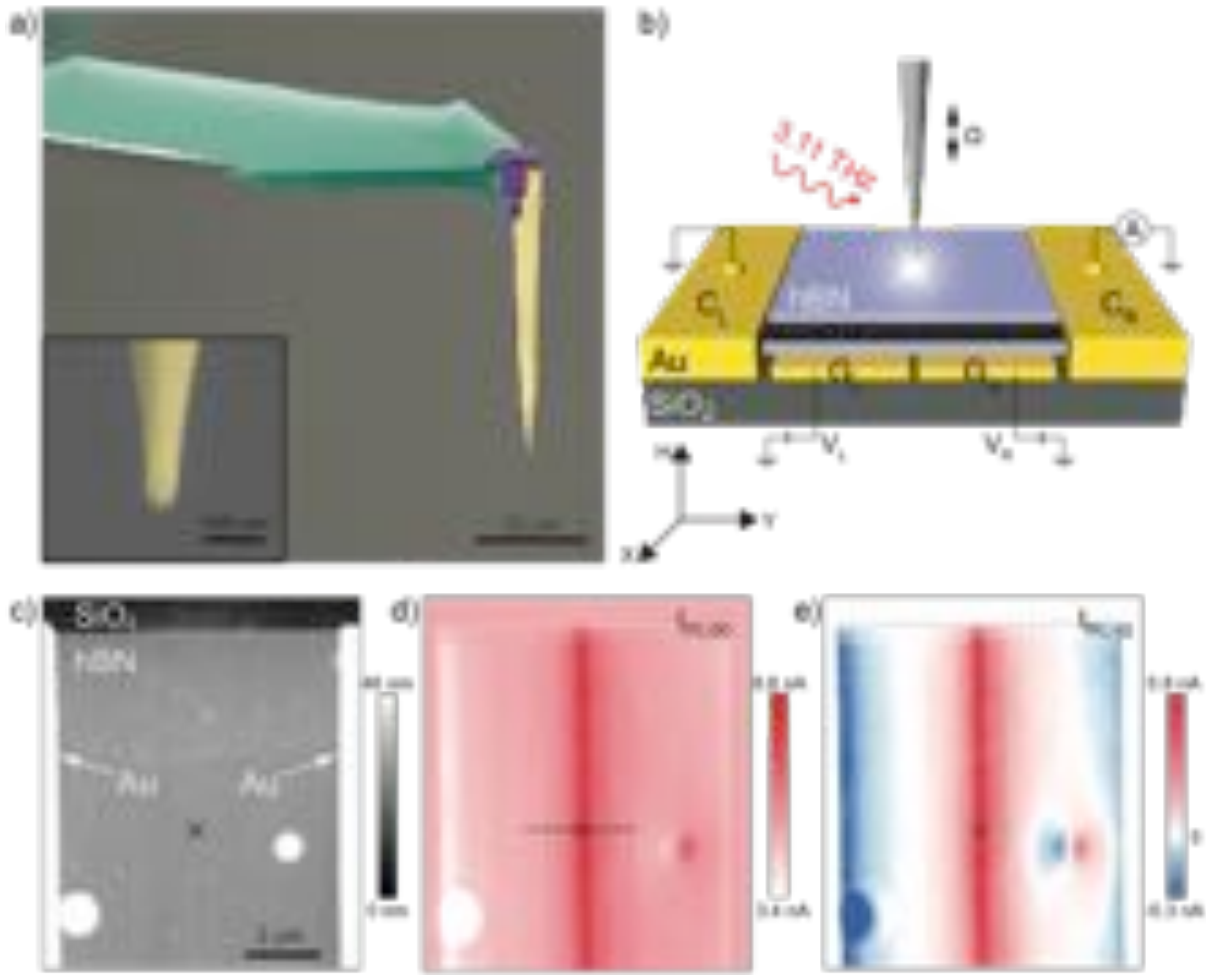

**Figure 1:** THz antenna tip and THz near-field detector: a) False color SEM image of a FIB fabricated THz antenna tip showing Si cantilever (green), focused ion beam deposited Pt (purple), and the Pt/Ir antenna tip (yellow). b) Schematics of the THz near-field detector. The laser illuminated antenna probe concentrates the light in the near-field region of the tip apex. The near-field induced photocurrent in the hBN-graphene-hBN heterostructure is detected through the two lateral contacts $C_L$ and $C_R$. Applying voltages $V_L$ and $V_R$ to the two backgates $G_L$ and $G_R$ allow to separately control the carrier concentration in the graphene to the left and to the right of the gap between them. c) AFM topography image of the THz near-field detector. d) + e) Images of direct (photo-)current (DC) $I_{PC,DC}$ and photocurrent recorded at frequency 1Ω $I_{PC,1\Omega}$. The white/gray dashed horizontal lines marks the edge of the graphene device. The black cross identifies the position of the measured approach curves shown in Fig. 2. The horizontal dashed black line marks the line profiles in Fig. 3.

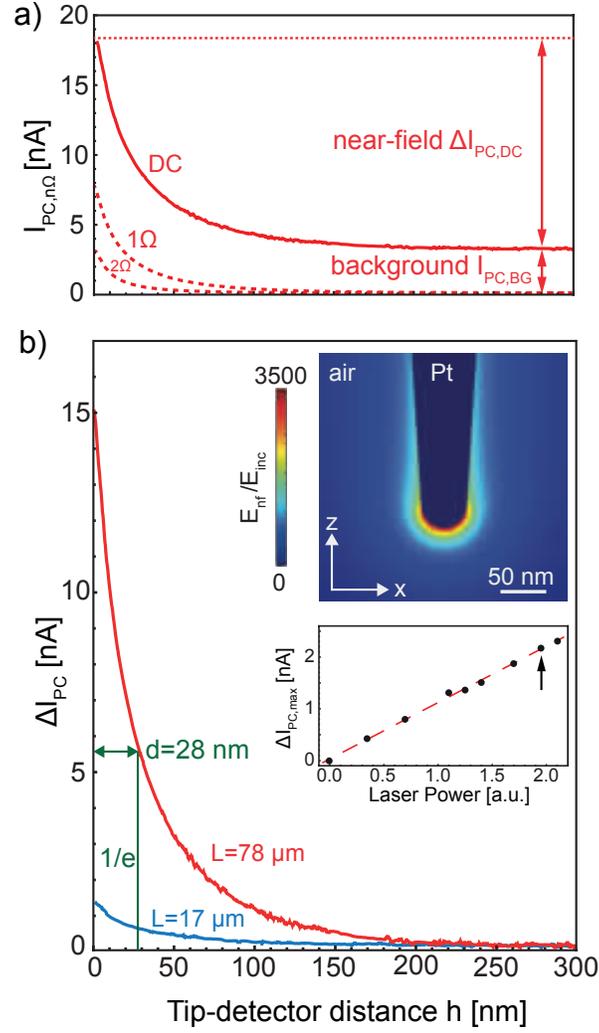

**Figure 2:** Photocurrent as a function of tip-detector distance. a) DC photocurrent $I_{PC,DC}$ and demodulated photocurrent $I_{PC,1\Omega}$ and $I_{PC,2\Omega}$. b) DC photocurrent after subtraction of background $\Delta I_{PC} = I_{PC,DC} - I_{PC,BG}$ for tips of lengths L = 78 $\mu$m (red) and 17 $\mu$m (blue). The upper inset shows the numerically calculated electric field distribution around the apex of a 78 $\mu$m long antenna tip. The lower inset shows the measured linear dependence of the photocurrent $\Delta I_{PC}$ on the THz laser illumination power (black dots), and a linear least-squares fit to the data (red dashed line). The arrow marks the power applied in the experiment.

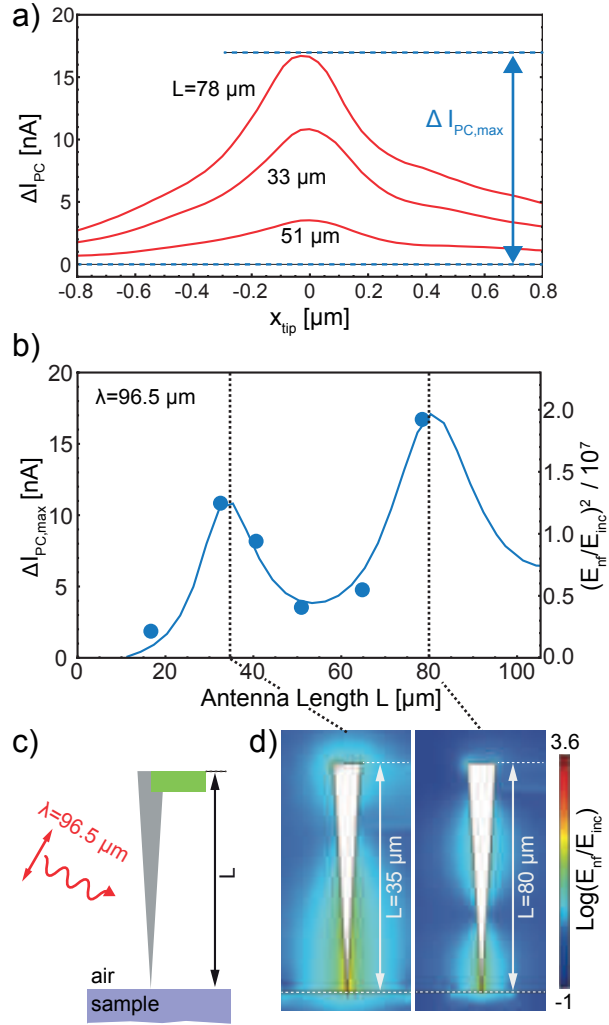

**Figure 3:** Evaluation of signal strength for different THz antenna tips. a) Photocurrent $\Delta I_{PC}$ line profiles for antenna tips with length 33 µm, 51 µm, and 78 µm. b) Maximum photocurrent $\Delta I_{PC,max}$ as a function of antenna length (blue dots) compared to numerical simulation (blue solid line). The vertical axes of the numerical simulation was manually adjusted such that best agreement to the experimental data points was obtained. c) Sketch of numerically simulated geometry (see more detail in Fig. 4a C) showing the tip (gray), the silicon cantilever (green), and the detector device (purple). d) False color image of the logarithm of the electric field enhancement of tips of length L = 35 µm and 80 µm, showing the first and second fundamental antenna resonance.

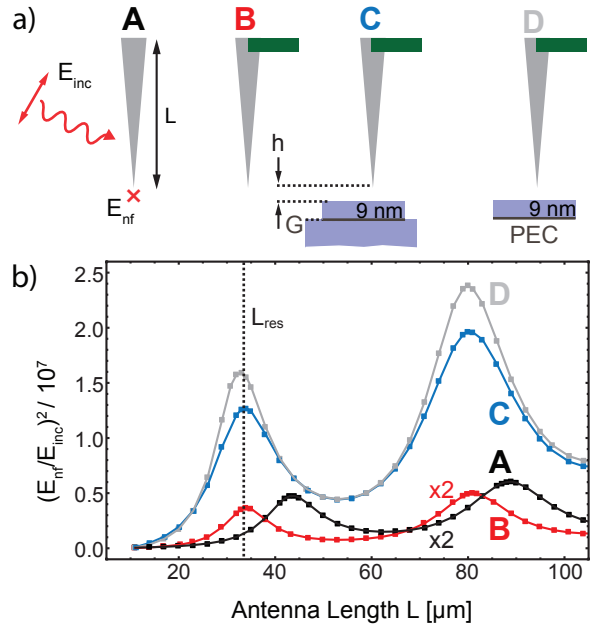

**Figure 4:** Evaluation of peak position of fundamental antenna resonances. a) Antenna geometries considered in the simulation. A: conical antenna tip. B: antenna tip with Si cantilever. C: as in B but with detector device below (9 nm hBN-graphene-bulk hBN) tip apex. D: as in C, replacing graphene with a PEC. b) Simulated antenna spectra for geometries A – D depicted in Fig. 4a.

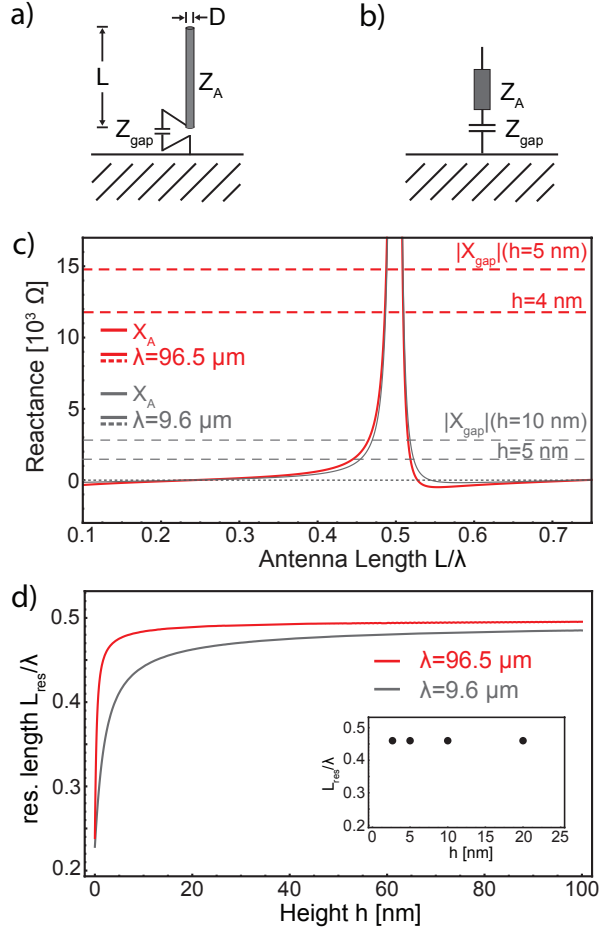

**Figure 5:** a, b) RF circuit model of a linear wire antenna above ground with input impedance $Z_{in}$, antenna impedance $Z_A$, and gap impedance $Z_{gap}$. c) Antenna reactance $X_A$ (solid lines) and gap reactance $X_{gap}$ (dashed lines) as a function of antenna length L for wavelength $\lambda = 96.5\ \mu m$ (red) and $\lambda = 9.6\ \mu m$ (black). d) Antenna resonance length $L_{res}$ normalized to the excitation wavelength $\lambda$ as a function of gap width h. The inset shows a numerical calculation of the resonance length $L_{res}$ for a THz antenna tip above a PEC.